\newif\ifsubmission
\submissionfalse
\ifsubmission
  \documentclass[sigconf,review]{acmart}
  \setcopyright{none}
  \acmConference[IWSiB 2027]{International Workshop on Software-intensive
    Business}{April 25--May 1, 2027}{Dublin, Ireland}
\else
  \documentclass[sigconf,nonacm]{acmart}
\fi
\usepackage{array}

\begin{document}

\title{A Carbon-Aware Quantum Computing Framework for LCA-Driven Sustainability in Quantum Cloud Services}

\author{Muhammad Umar}
\email{Muhammad.Umar@lut.fi}
\orcid{0009-0009-7251-7520}
\correspondingauthor
\affiliation{%
  \institution{LUT University}
  \city{Lappeenranta}
  \country{Finland}
}

\author{Nauman Arshad}
\email{Nauman.Arshad@lut.fi}
\affiliation{%
  \institution{LUT University}
  \city{Lappeenranta}
  \country{Finland}
}

\author{Azeem Akbar}
\email{Azeem.Akbar@lut.fi}
\affiliation{%
  \institution{LUT University}
  \city{Lappeenranta}
  \country{Finland}
}

\author{Arif Ali Khan}
\email{arif.khan@oulu.fi}
\affiliation{%
  \institution{University of Oulu}
  \city{Oulu}
  \country{Finland}
}

\renewcommand{\shortauthors}{Umar et al.}

\begin{abstract}
\textbf{Context:} Quantum computing's environmental footprint remains poorly understood relative to classical infrastructure, and as quantum computing moves toward cloud delivery, Quantum Cloud Service (QCS) providers lack actionable guidance beyond platform-level carbon-accounting frameworks. \textbf{Objective:} This study extends the carbon-aware quantum computing (CQC) framework from a platform-level to a service-level model that translates empirical life cycle assessment (LCA) findings of a superconducting quantum computer into guidance for QCS providers. \textbf{Method:} We modeled the CQC framework via service-level embodied-carbon allocation, load-independent and load-proportional operational decomposition, and a workload-resolved application offset on the basis of results acquired through a cradle-to-grave LCA of a superconducting quantum platform. \textbf{Results:} The five-year footprint is 583~t~CO\textsubscript{2}e (GKP) and 10{,}570~t (surface-code), dominated by embodied carbon (77.3--85.2\%), with operational-embodied parity not reached until 17.0--28.7 years versus 2.7 years for classical comparators. This reorders provider levers: utilization yields the largest gain (19.7$\times$), followed by service life extension (59.9\%) and electricity supply (6.5$\times$), while operational efficiency and renewable procurement offer limited leverage. \textbf{Conclusion:} Superconducting quantum computers are structurally embodied-carbon-dominated, inverting classical sustainability intuition and motivating direct power measurement and cross-architecture validation as quantum infrastructure scales.
\end{abstract}

\keywords{Quantum computing, quantum cloud services, carbon-aware computing, life cycle assessment, embodied carbon, operational carbon, sustainability, superconducting quantum computing}

\maketitle

\section{Introduction}

In contemporary computing technologies, Quantum Computing (QC) has emerged as a relatively new paradigm that complements High-Performance Computing (HPC), with the potential to solve problems that are intractable on classical infrastructure \cite{Cordier2025}. Though an emerging research field, it is yet unclear how the transformative impacts of QC will translate into the industrial and business infrastructure of tomorrow. Furthermore, environmental considerations incorporating climate change, carbon footprint, and Sustainable Development (SD) as a means to mitigate these also constitute potential challenges as QC progresses towards the Fault-Tolerant Computing (FTC) era \cite{Naumann2011}. Therefore, it is imperative to navigate the future QC infrastructures in terms of energy consumption and to ensure the technology's adherence to the United Nations Sustainable Development Goals (SDGs).

The environmental impact of quantum computers remains relatively underexplored compared to that of classical computers \cite{Cordier2025}. This may be attributed to the current Noisy Intermediate-Scale Quantum (NISQ) era and because Quantum Error Correction (QEC) is required for solving industrial-scale problems. Classical computing systems on the other hand have been extensively studied for their environmental impacts through life cycle assessment (LCA). LCA is a comprehensive method that identifies environmental burdens and informs decision making while evaluating systems from material procurement to end of life. This method has been widely applied across classical computing contexts. Research studies have examined data center infrastructure beyond operational energy use \cite{Shah2011}, energy variations across processor types \cite{Bol2011}, and supercomputer designs where cooling stages drive air acidification while cooling manufacturing dominates ozone depletion and cancer risk impacts \cite{McDonnell2013}.  Other work has compared desktops with all-in-ones \cite{Subramanian2017}, shown lower emissions from thin-client server-based computing versus desktops \cite{Maga2012},  and demonstrated the environmental benefits of server-supported single-board computers over traditional desktop PCs \cite{Loubet2023}. 

On the contrary, limited research literature on the environmental impacts estimation of quantum computers' usage and production is available. \cite{Billat2024} and \cite{Cordier2025} constitute some of the LCA studies conducted on quantum computers so far. \cite{Billat2024} conducted an LCA to estimate the environmental impacts of a superconducting quantum computer with identification of cryostat as the subsystem with highest impact contribution during the production phase. \cite{Cordier2025} conducted a comparative LCA study to assess the potential environmental impact of a quantum vs a classical computer. The results of this study emphasized QEC hardware to have a substantial environmental impact due to the large number of electronic components needed to achieve the required number of logical qubits for a fault-tolerant quantum computer.

The few existing LCAs of quantum computing facilities also lack a framework through which their findings could be translated into guidance for software-intensive businesses built on quantum computing. The present work therefore aims to bridge this gap by conducting a life cycle assessment (LCA) of a quantum computer to quantify its environmental burdens across key life-cycle stages, and to use these findings as an empirical foundation for a framework that guides Quantum Cloud Service (QCS) providers toward more sustainable operational and infrastructural practices. As quantum computing increasingly moves toward cloud-based delivery, QCS providers are positioned to play a central role in shaping the technology's environmental trajectory. Equipping them with evidence-based guidance rather than assumptions extrapolated from classical computing practices is therefore essential to steering QC development along a sustainable path. Accordingly, this study is guided by the following research questions:

\begin{itemize}
\item \textbf{RQ1:} What are the environmental burdens associated with a quantum computer across its life cycle stages, from production through operational use to end of life?
\item \textbf{RQ2:} Which subsystems or lifecycle stages of the quantum computer contribute most significantly to its overall environmental impact?
\item \textbf{RQ3:} How can the empirical findings from the conducted LCA be translated into actionable guidance for Quantum Cloud Service providers to support more sustainable quantum computing practices?
\end{itemize}

The novelty of this work is twofold. First, it extends the limited body of quantum computing LCA literature, currently represented by studies such as \cite{Billat2024} and \cite{Cordier2025}, by providing additional empirical data on the environmental burdens of quantum computing systems. Second, and more significantly, it moves beyond impact quantification to translate LCA findings into a practical, actionable framework tailored specifically for QCS providers. While existing studies have quantified impacts at the subsystem level or compared quantum against classical architectures, none have yet used such findings to inform sustainability-oriented decision-making frameworks for cloud-based quantum service delivery. By bridging empirical LCA results with practice-oriented recommendations, this study supports QCS providers in aligning their offerings with the SDGs, while establishing a methodological foundation for future work as quantum computing infrastructure continues to scale toward the fault-tolerant era.

\section{Methods}

This paper addresses the prevailing gap between LCA-derived environmental data for QC infrastructure and actionable sustainability guidance for Quantum Cloud Service (QCS) providers. The methodology proceeds in two parts. Section~\ref{sec:lca} specifies the life cycle assessment used to characterize a reference quantum computing platform empirically. Section~\ref{sec:cqc} then extends the carbon-aware quantum computing (CQC) framework of Arora and Kumar~\cite{Arora2025} from a platform-level accounting model to a service-level model, instantiated with the inventory data established in Section~\ref{sec:lca}.

\subsection{Research Design Overview}
\label{sec:design}

The two parts of the methodology map directly onto the paper's research questions. RQ1 and RQ2, concerning the environmental burdens of a quantum computer across its life cycle and the subsystems or stages that dominate those burdens, are addressed empirically through the life cycle assessment of Section~\ref{sec:lca}. RQ3, concerning how these empirical findings can be translated into actionable guidance for QCS providers, is addressed through the service-level extension of the CQC framework in Section~\ref{sec:cqc}.

\subsection{Life Cycle Assessment (LCA)}
\label{sec:lca}

\subsubsection{Goal and Scope}

The goal of this LCA study is to assess the potential environmental impact of quantum computing infrastructure. A superconducting quantum computing architecture is selected as the reference system. The scope of the assessment is to analyze how the environmental impact of a superconducting quantum computer evolves over its operational lifetime.

The modeled system reflects the four principal subsystems comprising a state-of-the-art cryogenic platform for superconducting quantum computing: the cryostat, the gas handling system (GHS), the compressor, and the control unit (CU). Subsystem modeling is based on technical specifications of the Bluefors first-generation system \cite{Bluefors2022}, with the cryostat modeled on the BlueFors LD400, the GHS on the GHS400, and the compressor on the CP1110. Gold-coated components within the cryostat are modeled with an assumed coating thickness of 1.5~\textmu m, and structural materials in the room housing the cryostat are assumed to be primarily aluminum, copper, or stainless steel, selected to minimize magnetic field interference and associated signal noise.

The functional unit adopted for this assessment is one platform-hour of the modeled superconducting quantum processing unit (QPU) under the reference 100-logical-qubit configuration described in Section~\ref{sec:lci_qec}. The system boundary is cradle-to-grave, encompassing production, delivery, use, and end-of-life phases as illustrated in Figure~\ref{fig:lci_phases}. Results expressed per delivered quantum job and per shot, together with the allocation rule by which the platform-hour functional unit is apportioned across jobs, are defined subsequently as part of the service-level extension in Section~\ref{sec:cqc}; downstream application-specific abatement is treated there as an optional credit outside the present system boundary.

\subsubsection{Life Cycle Inventory}
\label{sec:lci_qec}

\begin{figure*}[t]
    \centering
    \includegraphics[width=\textwidth]{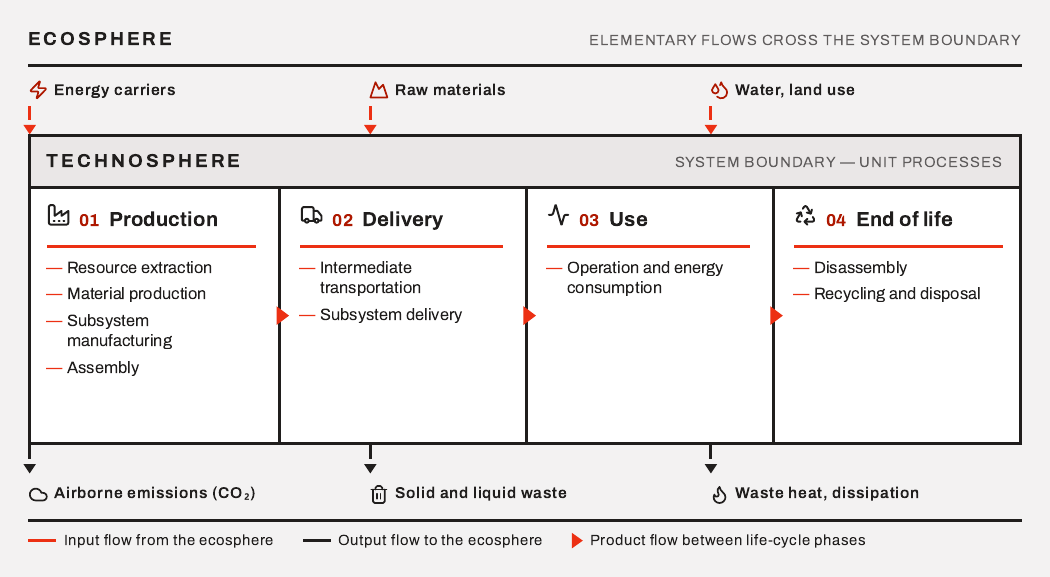}
    \caption{System boundary and life-cycle flow diagram, showing the four technosphere phases (production, delivery, use, end of life) and their elementary exchanges with the ecosphere.}
    \Description{A flow diagram illustrating the system boundary and four life-cycle phases: production, delivery, use, and end of life, with arrows showing exchanges with the ecosphere.}
    \label{fig:lci_phases}
\end{figure*}

The life cycle inventory (LCI) is built using the ecoinvent database (version 3.9, allocation cut-off by classification) \cite{Wernet2016}, which supplies background data representing the broader network of human activities connected to the system under study. Foreground data, describing the quantum computing system technically, are derived primarily from on-site data collection, technical documentation, and system drawings. The inventory is structured around four life cycle phases: production, delivery, use, and end-of-life, as illustrated in Figure~\ref{fig:lci_phases}.

\paragraph{Production}
The production phase encompasses resource extraction, material production, subsystem manufacturing, assembly, and intermediate transportation. Following resource extraction and material production, manufacturing processes are modeled to shape the constituent parts. For the intermediate manufacturing step occurring between material production and machining, the raw material mass is used as the modeling basis rather than the final, post-machining component mass, in order to account for material losses during fabrication.

\paragraph{Delivery}
The delivery phase accounts for the transportation of all subsystems from their respective sourcing locations to the point of assembly. The cryostat and its support frame, the GHS, and the CU are assumed to be sourced from Finland. The compressor is assumed to be sourced from the United States, while its associated liquid nitrogen tank is sourced from France. Quantum error correction (QEC) electronics and their associated cabling, with cable quantities scaled proportionally to the size of the QEC system, are assumed to be sourced from the United States.

\paragraph{Use}
The use phase models the system's operational energy and material consumption, excluding sleep or off-mode states, so that the system is assumed to be continuously active. The use phase also incorporates a desktop computer used for data management and computation initiation. Operationally, the system requires electricity and periodic nitrogen refueling, estimated at approximately 10~L per week per compressor; lost nitrogen is modeled as an elementary flow emitted to air. Helium consumption is accounted for exclusively within the production phase, since the closed-loop cooling circuit is assumed to experience no losses or refueling requirements during use.

Power consumption during the use phase scales with the number of qubits, incorporating both scaling and multiplexing factors. Electronics energy consumption is estimated using component-level data reported by Lachance-Quirion et al.~\cite{LachanceQuirion2024}: cryostat electronics require approximately 36~W per logical qubit, and electronic racks (mainly oscillator cards) require approximately 209~W per logical qubit. A single GHS operating with two compressors requires approximately 1.8~kW, while each compressor independently requires approximately 10.7~kW. Power consumption of the CU is assumed negligible, based on typical daily laboratory usage patterns.

\paragraph{End of Life}
End-of-life modeling follows the cut-off approach, under which waste management responsibility is assigned to the producer. Recyclable products are treated as burden-free at end-of-life, since the environmental burden of recycling processes is instead allocated to the subsequent production cycle that uses the recycled material \cite{EcoinventSystemModels2024}.

\paragraph{Quantum Error Correction}
To reflect a fault-tolerant operating regime, the single-logical-qubit QEC setup demonstrated by Lachance-Quirion et al.~\cite{LachanceQuirion2024} is scaled to 100 logical qubits. Seven physical qubits are assumed per logical qubit, yielding an overhead factor of $O = 7$ based on Steane's code \cite{Cai2021}. A multiplexing factor of $M = 4$ is applied to account for electronic equipment and cabling shared across multiple qubits \cite{Shi2023}. The combined scaling factor, $O/M = 7/4$, implies that 175 times the original single-qubit setup is required to reach 100 logical qubits. To accommodate the corresponding increase in electronics and cabling, the cryostat, support frame, and compressor are each scaled by a factor of six, with three GHS units and one CU determined to be sufficient to operate the resulting six cryostats.

The resulting system has an estimated total power consumption of approximately 112.6~kW, dominated by the compressors (64~kW) and QEC electronics (43~kW). This elevated energy demand, relative to conventional (non-error-corrected) superconducting quantum processors, is attributed primarily to the additional hardware required to implement QEC across 100 logical qubits. The modeled system performs autonomous QEC of bosonic Gottesman-Kitaev-Preskill (GKP) states, and therefore does not require supplementary classical decoding hardware in the baseline scenario.

\subsubsection{Sensitivity Analysis}
\label{sec:sensitivity}

Given that the efficiency of QEC and the associated multiplexing factor are strongly dependent on experimental noise characteristics and hardware architecture, these parameters are treated as key sources of uncertainty and are examined through a dedicated sensitivity analysis. An alternative scenario, denoted Scenario~A$'$, scales the Lachance-Quirion et al. \cite{LachanceQuirion2024} setup by a factor of 2500 (rather than 175), reflecting assumptions more representative of surface-code architectures, which typically require higher physical-to-logical qubit ratios than bosonic GKP-based approaches. Because bosonic GKP and surface codes follow distinct scaling rules, the multiplexing and cryostat scaling factors are adjusted accordingly between scenarios.

Two physical-to-logical qubit ratios are examined: a conservative upper-bound ratio of 1000:1, accounting for multiple layers of encoding (Scenario~A$'$), and a less conservative ratio of 100:1, assuming a single encoding layer is sufficient for fault-tolerant operation (Scenario~A). Scenario~A includes the classical hardware required for syndrome decoding during QEC, whereas Scenario~A$'$ excludes this decoding overhead due to the current lack of sufficient data for detailed modeling, a limitation that is revisited in the discussion of results.

\subsubsection{Life Cycle Impact Assessment Methodology}

Characterization of elementary flows is performed using the IMPACT World+ methodology (Expert version 2.0.1), which resolves inventory flows into 21 midpoint damage indicators grouped under two areas of protection: human health and ecosystems. Climate change is assessed separately using the IPCC 2021 characterization factors \cite{IPCC2021} and is expressed in tonnes of CO\textsubscript{2}-equivalent (t~CO\textsubscript{2}-eq.), enabling aggregation of heterogeneous greenhouse gas emissions according to their global warming potential. The ecosystems indicator is expressed as the potentially disappeared fraction of species, in PDF$\cdot$m$^2\cdot$yr, while the human health indicator is expressed in disability-adjusted life years (DALYs), reflecting years of life lost to premature death, disability, or disease. Life cycle inventory management and characterization calculations are performed in SimaPro~9 \cite{PReSustainability2024}.

\subsection{Extending the Carbon-Aware Quantum Computing (CQC) Framework}
\label{sec:cqc}

\subsubsection{Baseline Formulation}

Arora and Kumar~\cite{Arora2025} formalized the total carbon footprint of a quantum computing platform as the sum of two life-cycle burdens less an application-derived credit:

\begin{equation}
CO_2e_{\text{total}} = CO_2e_{\text{embodied}} + CO_2e_{\text{operational}} - CO_2e_{\text{application}}
\label{eq:cqc}
\end{equation}

Here $CO_2e_{\text{embodied}}$ accounts for materials extraction, manufacturing, and transport; $CO_2e_{\text{operational}}$ accounts for energy, water, and cooling consumed during use; and $CO_2e_{\text{application}}$ represents a negative offset attributable to sustainability-oriented workloads executed on the platform. The first two terms characterize \textit{sustainability for QC}, while the third characterizes \textit{QC for sustainability} \cite{Arora2025}.

\subsubsection{Limitations for Service-Level Decision-Making}

Equation~\ref{eq:cqc} is defined over the life cycle of a single platform. This level of detail, though appropriate for hardware designers and manufacturers, does not correspond to the decision boundary occupied by a QCS provider, for three reasons.

First, a cloud-delivered QPU is a \textit{multi-tenant} resource: a single physical system serves many independent users, so a per-platform footprint cannot be attributed to any individual consumer of the service without an explicit allocation rule. Second, cryogenic quantum systems exhibit a strongly \textit{load-independent} power profile: the cryostat, gas handling system, and compressors must be sustained at operating temperatures irrespective of whether circuits are executing, so operational carbon does not scale linearly with computational demand as it does for classical cloud infrastructure. Third, QCS providers exercise control over a distinct set of levers incorporating scheduling, batching, queue management, siting, and hardware refresh policy, none of which are visible in a formulation whose only variables are hardware composition and cumulative energy draw.

We therefore extend Equation~\ref{eq:cqc} along three axes: (i) a service-level allocation of the platform-hour functional unit established in Section~\ref{sec:lca}; (ii) a decomposition of operational carbon into load-independent and load-proportional components under a time-varying grid carbon intensity; and (iii) a workload-resolved formulation of the application offset.

\subsubsection{Service-Level Allocation of Embodied Carbon}

Building on the platform-hour functional unit and cradle-to-grave system boundary established in Section~\ref{sec:lca}, let a QCS platform have an operational lifetime $T_{\text{life}}$ (hours) and let a submitted quantum job $j$ occupy the QPU for an allocated execution window $\tau_j$ (hours), comprising $n_j$ shots. We define the allocation factor over the computation the platform delivers across its life:

\begin{equation}
\alpha_j = \frac{\tau_j}{\sum_{k} \tau_k} = \frac{\tau_j}{U\,T_{\text{life}}}
\label{eq:alloc}
\end{equation}

where the sum runs over all jobs served within $T_{\text{life}}$ and $U$ is the platform utilization (Equation~\ref{eq:util}) over the same period. The rule is idle-inclusive: hours in which the platform is held at operating conditions without executing jobs are charged to the computation it delivers, so the allocation factors of all jobs sum to one and the whole embodied burden is attributed. Allocating over wall-clock life instead, as $\tau_j / T_{\text{life}}$, would leave a fraction $1-U$ of the burden attributed to no job. The embodied carbon attributable to job $j$ can then be written as:

\begin{equation}
C^{j}_{\text{embodied}} = \alpha_j \cdot CO_2e_{\text{embodied}}
\label{eq:emb_job}
\end{equation}

where $CO_2e_{\text{embodied}}$ is obtained from the production, delivery, and end-of-life phases of the LCI (Section~\ref{sec:lca}). Equation~\ref{eq:emb_job} makes explicit a property that the platform-level formulation obscures: because $CO_2e_{\text{embodied}}$ is fixed at manufacture, the embodied carbon borne by each job is inversely proportional to the total volume of computation the platform delivers over its lifetime. Extending $T_{\text{life}}$ or increasing throughput reduces per-job embodied carbon without altering the hardware itself.

\subsubsection{Load-Independent and Load-Proportional Operational Carbon}

We decompose the platform's instantaneous power demand into a baseline term $P_{\text{base}}$, sustained continuously to maintain cryogenic and vacuum conditions, and a marginal term $P_{\text{marg}}$ drawn only during circuit execution:

\begin{equation}
P(t) = P_{\text{base}} + P_{\text{marg}} \cdot \mathbb{I}_{\text{exec}}(t)
\label{eq:power_split}
\end{equation}

where $\mathbb{I}_{\text{exec}}(t)$ is an indicator function equal to unity while a job is executing. Operational carbon over an accounting period $[0,T]$ follows by integrating against the time-varying carbon intensity $I(t)$ of the electricity supply (kg CO\textsubscript{2}e per kWh):

\begin{equation}
CO_2e_{\text{operational}} = \int_{0}^{T} \big[ P_{\text{base}} + P_{\text{marg}} \cdot \mathbb{I}_{\text{exec}}(t) \big] \, I(t) \, dt
\label{eq:op_integral}
\end{equation}

Two consequences follow directly, and both are actionable at the provider level rather than the hardware level. Where $P_{\text{base}} \gg P_{\text{marg}}$, the regime characteristic of superconducting architectures whose compressors and cryogenic subsystems dominate the power budget, operational carbon is governed principally by \textit{elapsed time under refrigeration} rather than by computational volume. Defining platform utilization over the accounting period as:

\begin{equation}
U = \frac{1}{T}\sum_{j} \tau_j
\label{eq:util}
\end{equation}

the operational carbon attributable to each unit of delivered computation falls approximately as $1/U$: an underutilized QPU distributes an essentially unchanged absolute footprint across fewer jobs. Consolidation, batching, and queue densification therefore act as carbon-reduction measures in their own right, independent of any improvement in hardware efficiency. Separately, retaining $I(t)$ inside the integral rather than applying an annual average admits carbon-aware scheduling: deferring latency-tolerant jobs toward low-intensity grid periods reduces $CO_2e_{\text{operational}}$ for a fixed workload, echoing the carbon-aware scheduler proposed as an open direction in~\cite{Arora2025}.

\subsubsection{Workload-Resolved Application Offset}

The third term of Equation~\ref{eq:cqc} is defined over the platform's application portfolio. In a QCS setting this portfolio is the provider's realized workload mix, which we resolve per workload class $i$:

\begin{equation}
CO_2e_{\text{application}} = \sum_{i} w_i \big( C^{\text{classical}}_{i} - C^{\text{QC}}_{i} \big)
\label{eq:app_offset}
\end{equation}

where $C^{\text{classical}}_{i}$ is the carbon cost of obtaining an equivalent result on classical HPC infrastructure, $C^{\text{QC}}_{i}$ is the carbon cost of executing workload class $i$ on the QPU as given by Equations~\ref{eq:emb_job} and~\ref{eq:op_integral}, and $w_i$ weights the realized downstream emissions reduction enabled by the result, for instance the abatement attributable to an optimized industrial process informed by a quantum simulation.

Quantifying $w_i$ requires domain expertise beyond the boundary of a hardware LCA, and no empirical basis for it exists in the current literature. We therefore retain the term structurally but set $CO_2e_{\text{application}} = 0$ throughout the present assessment, so that reported footprints constitute a conservative upper bound. Equation~\ref{eq:app_offset} defines the interface through which provider-specific workload data would populate this term in future work.

\subsubsection{Extended Formulation and Mapping to the Life Cycle Inventory}

Combining Equations~\ref{eq:emb_job},~\ref{eq:op_integral}, and~\ref{eq:app_offset}, the carbon footprint of a quantum job delivered as a cloud service is:

\begin{equation}
\begin{split}
C_{j} ={}&
\underbrace{\alpha_j \, CO_2e_{\text{embodied}}}_{\text{cradle-to-grave, amortized}}
+ \underbrace{\frac{\tau_j}{\sum_{k} \tau_k} \int_{0}^{T} P_{\text{base}}\, I(t)\, dt}_{\text{baseline, allocated}} \\
&+ \underbrace{\int_{\tau_j} P_{\text{marg}}\, I(t)\, dt}_{\text{API-to-shot, executed}}
- \underbrace{w_j \big( C^{\text{classical}}_{j} - C^{\text{QC}}_{j} \big)}_{\text{application credit}}
\end{split}
\label{eq:cqc_qcs}
\end{equation}

with the corresponding per-shot intensity $C_j / n_j$ serving as the reportable service-level metric. The baseline term follows the same idle-inclusive rule as $\alpha_j$, with the sum taken over the jobs executed in the accounting period $[0,T]$: the load-independent draw is shared among the jobs the platform serves, while the marginal draw is charged to the job that causes it. Summed over all jobs, Equation~\ref{eq:cqc_qcs} therefore recovers the platform's full embodied and operational footprint. Each term is addressable through decisions a QCS provider actually makes: hardware lifetime and refresh policy govern the amortization of $CO_2e_{\text{embodied}}$; utilization governs both allocated terms; batching and siting govern the operational integrals; and workload portfolio composition governs the application credit.

Table~\ref{tab:cqc_mapping} states the correspondence between the LCI phases specified in Section~\ref{sec:lca} and the terms of Equation~\ref{eq:cqc_qcs}, together with the provider-level lever each term exposes.

\begin{table}[h]
\caption{Correspondence between LCI phases, extended CQC terms, and QCS provider levers.}
\label{tab:cqc_mapping}
\small
\begin{tabular}{>{\raggedright\arraybackslash}p{1.7cm}>{\raggedright\arraybackslash}p{2.2cm}>{\raggedright\arraybackslash}p{2.6cm}}
\toprule
LCI phase & CQC term & Provider lever \\
\midrule
Production, Delivery, End-of-Life & $\alpha_j \, CO_2e_{\text{embodied}}$ & Lifetime extension, modular refresh, component reuse \\
Use & $\int P\,I(t)\,dt$ & Utilization, batching, carbon-aware scheduling, siting \\
Not assessed & $CO_2e_{\text{application}}$ & Workload portfolio composition \\
\bottomrule
\end{tabular}
\end{table}

\section{Results and Discussion}
\label{sec:results}

This section instantiates the extended CQC formulation of Section~\ref{sec:cqc} with the life cycle inventory specified in Section~\ref{sec:lca}. We first decompose the assessed climate change impact into the embodied and operational terms of Equation~\ref{eq:cqc} and establish the point at which their contributions reach parity. We then evaluate the three provider-level levers exposed by the service-level formulation, namely utilization, service life, and electricity supply, and close by examining the status of the application-centric offset term.

All results are reported for the climate change indicator, characterized using IPCC 2021 GWP100 factors and expressed in tonnes of CO\textsubscript{2}-equivalent. The carbon intensity of the modeled electricity supply, back-calculated from the use-phase inventory and the installed power of Scenario~A, is 0.0268\,kg\,CO\textsubscript{2}e per kWh, reflecting the low-carbon hydroelectric market process used throughout the inventory. Unless stated otherwise, the service life is taken as $T_{\text{life}} = 43{,}800$\,h (five years) and utilization as $U = 0.50$.

\subsection{Embodied and Operational Decomposition}
\label{sec:res_decomp}

Table~\ref{tab:decomp} reports the decomposition of each assessed scenario into the two life-cycle terms of Equation~\ref{eq:cqc}. The production, delivery, and end-of-life phases are aggregated into $CO_2e_{\text{embodied}}$; the use phase yields the operational accrual rate.

\begin{table}[t]
\caption{Decomposition of the assessed climate change impact into the embodied and operational terms of the CQC framework. Parity time $t^{*}$ denotes the operating duration at which cumulative operational carbon equals embodied carbon.}
\label{tab:decomp}
\small
\setlength{\tabcolsep}{4pt}
\begin{tabular}{lrrrr}
\toprule
Scenario & Embodied & Rate & $t^{*}$ & $t^{*}$ \\
         & (t CO\textsubscript{2}e) & (t/h) & (h) & (yr) \\
\midrule
A (GKP QEC)        & 450.97    & 0.00302 & 149{,}313 & 17.0 \\
A$'$ (surface code) & 9{,}000.45 & 0.03583 & 251{,}211 & 28.7 \\
B (classical HPC)   & 11{,}711.08 & 0.48809 & 23{,}994  & 2.7 \\
B$'$ (TOP500 class) & 4{,}391.42 & 0.18303 & 23{,}993  & 2.7 \\
\bottomrule
\end{tabular}
\end{table}

The embodied term dominates both quantum scenarios by a wide margin. In Scenario~A, production alone accounts for 447.74\,t\,CO\textsubscript{2}e of a 450.97\,t embodied total, with delivery and end-of-life contributing 1.62\,t and 1.61\,t respectively. Scaling from the bosonic GKP configuration to the surface-code representative configuration of Scenario~A$'$ raises embodied carbon by a factor of 20.0 and the operational rate by a factor of 11.9, confirming that the hardware multiplication required by less efficient error-correcting codes propagates more strongly into the production phase than into operation.

\subsection{Embodied and Operational Parity}
\label{sec:res_parity}

\begin{figure}[htbp]
\centering
\includegraphics[width=\columnwidth]{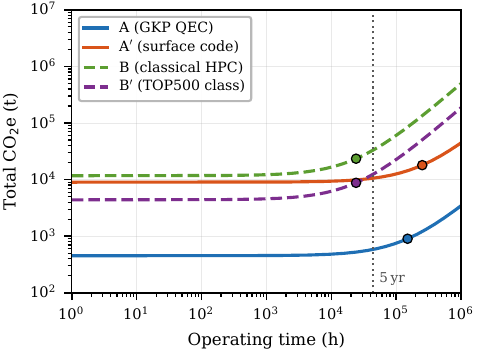}
\caption{Life-cycle carbon trajectory against operating time. Markers denote the embodied and operational parity point $t^{*}$; the dotted line marks the five-year service life assumed throughout.}
\Description{Log-log plot of total CO2-equivalent against operating hours for four scenarios. The two quantum scenarios remain flat far longer than the two classical comparators before rising.}
\label{fig:lifecycle}
\end{figure}

Figure~\ref{fig:lifecycle} plots the life-cycle trajectory $C_{\text{total}}(t) = C_{\text{embodied}} + \dot{c}_{\text{op}} t$ for all four scenarios, with markers indicating the parity time $t^{*}$ at which cumulative operational carbon equals the embodied burden.

The separation between quantum and classical platforms is pronounced. Both classical comparators reach parity at approximately 24{,}000\,h, that is 2.7 years, comfortably inside a conventional hardware refresh cycle. The quantum scenarios reach parity only at 17.0 years (Scenario~A) and 28.7 years (Scenario~A$'$), well beyond any plausible service life for a first-generation error-corrected platform. Over any realistic deployment horizon, therefore, an error-corrected superconducting quantum computer is structurally an embodied-carbon-dominated system, whereas classical high-performance computing infrastructure is operational-carbon-dominated.

\begin{figure}[htbp]
\centering
\includegraphics[width=\columnwidth]{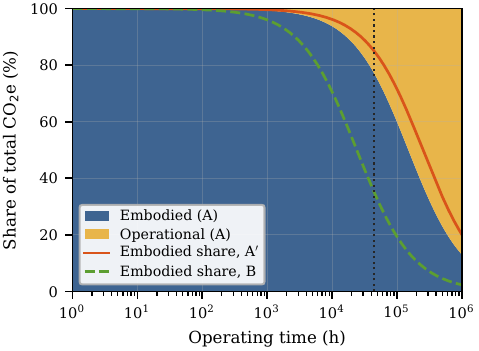}
\caption{Proportional contribution of the embodied and operational terms over time. Shaded regions correspond to Scenario~A; overlaid curves give the embodied share for Scenario~A$'$ and the classical comparator.}
\Description{Stacked area chart showing embodied share declining and operational share rising with operating hours, with the classical scenario crossing fifty percent much earlier than either quantum scenario.}
\label{fig:share}
\end{figure}

Figure~\ref{fig:share} makes the same point in proportional terms. At the five-year mark, operational carbon accounts for 22.7\% of the total footprint in Scenario~A and 14.8\% in Scenario~A$'$, against 64.6\% for the classical Scenario~B.

This finding carries a direct methodological consequence. Sustainability assessments that characterize quantum computing solely through operational energy demand, which remains the prevailing practice in the energy-efficiency literature, capture less than a quarter of the actual life-cycle burden of an error-corrected platform. It also inverts the intuition transferred from classical cloud infrastructure, where operational efficiency measures are the dominant lever. For quantum cloud services, the corresponding leverage lies in how thoroughly a fixed manufacturing burden is amortized.

\subsection{Load-Independent Power Dominance}
\label{sec:res_power}

Applying the decomposition of Equation~\ref{eq:power_split} to the use-phase inventory quantifies the asymmetry that motivates the service-level extension. Compressors (64.2\,kW across six units) and gas handling systems (5.4\,kW across three units) are sustained continuously irrespective of computational load. The cryostat and QEC electronics (42.96\,kW) are split by $\phi$, the load-proportional fraction, which we set to 0.20 on the basis that autonomous GKP correction operates continuously to preserve logical qubit states and therefore contributes principally to the baseline.

Under this split, Scenario~A yields $P_{\text{base}} = 104.0$\,kW against $P_{\text{marg}} = 8.6$\,kW, a ratio of 12.1:1. Scenario~A$'$ yields 1{,}091.4\,kW against 243.8\,kW, a ratio of 4.5:1. The classical comparators, whose equivalent power is back-calculated from their use-phase rate and split at $\phi_{cl} = 0.60$ to reflect the wider dynamic range of conventional server hardware, invert this relationship to 0.67:1.

The consequence is that operational carbon in a superconducting quantum platform is governed principally by elapsed time under refrigeration rather than by computational volume. An idle QPU held at 20\,mK draws close to the same power as a fully loaded one. This is not a marginal correction to the classical picture but a qualitative reversal of it, and it is invisible in a platform-level accounting model that reports only cumulative energy.

\subsection{Utilization as a Provider-Level Lever}
\label{sec:res_util}

\begin{figure}[htbp]
\centering
\includegraphics[width=\columnwidth]{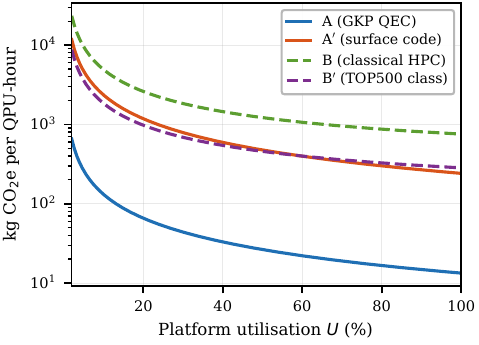}
\caption{Carbon intensity of delivered service against platform utilization, at a five-year service life. The near-inverse scaling reflects the dominance of the load-independent baseline.}
\Description{Semi-log plot of kilograms CO2-equivalent per QPU-hour against utilization percentage, declining steeply for all four scenarios.}
\label{fig:utilization}
\end{figure}

Because both the amortized embodied term and the dominant baseline power term are effectively fixed per unit of wall-clock time, the carbon intensity of delivered computation scales approximately as $1/U$. Figure~\ref{fig:utilization} quantifies this relationship.

For Scenario~A, carbon intensity falls from 261.9\,kg\,CO\textsubscript{2}e per QPU-hour at 5\% utilization to 13.3\,kg at full utilization, a reduction of a factor of 19.7 achieved without any change to the hardware, the error-correcting code, or the electricity supply. Scenario~A$'$ exhibits the same scaling, from 4{,}702\,kg to 241\,kg per QPU-hour.

Utilization is therefore a first-order sustainability lever for quantum cloud service providers, and one they already control through queue management, job batching, and workload consolidation across a fleet. It is also, notably, a lever that aligns with commercial incentives rather than opposing them, since the measures that reduce carbon intensity per delivered job are the same measures that improve return on a capital-intensive asset. This alignment is worth emphasizing because sustainability interventions in computing more commonly involve a performance or cost penalty.

\subsection{Service Life Extension}
\label{sec:res_life}

\begin{figure}[htbp]
\centering
\includegraphics[width=\columnwidth]{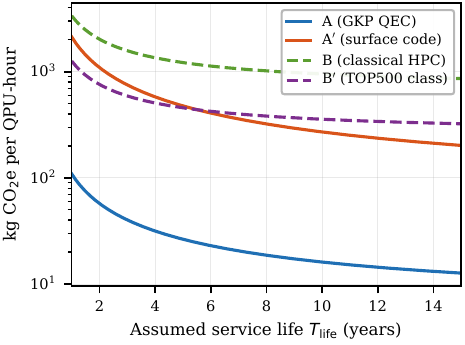}
\caption{Service-level carbon intensity against assumed hardware service life, at 50\% utilization. The steep initial gradient reflects the embodied-dominated regime of the quantum scenarios.}
\Description{Semi-log plot of kilograms CO2-equivalent per QPU-hour against service life in years, declining steeply between one and five years and flattening thereafter.}
\label{fig:lifetime}
\end{figure}

The allocation factor of Equation~\ref{eq:alloc} implies that per-job embodied carbon falls inversely with the total computation a platform delivers across its life. Figure~\ref{fig:lifetime} evaluates this lever across service lives from one to fifteen years.

Extending the assumed service life of Scenario~A from three to ten years reduces carbon intensity from 40.1 to 16.1\,kg\,CO\textsubscript{2}e per QPU-hour, a 59.9\% reduction. The gradient is steepest in the first five years and flattens thereafter, indicating that early-life replacement is disproportionately costly in carbon terms and that the marginal benefit of extension diminishes once the platform passes roughly a decade of service.

The magnitude of this lever is a direct consequence of the embodied dominance established in Section~\ref{sec:res_parity}. In an operational-carbon-dominated system such as the classical comparators, lifetime extension yields a much shallower response, and can be counterproductive where newer hardware offers materially better energy efficiency per unit of computation. For quantum platforms in the current regime, no such tension arises: because operational carbon is a minority contributor across any plausible service life, the embodied saving from retention is not readily offset by efficiency gains from replacement. This supports modular hardware design and component-level refresh over whole-system replacement, consistent with the disposal-phase recommendations of the CQC framework~\cite{Arora2025}.

\subsection{Electricity Supply and Carbon-Aware Scheduling}
\label{sec:res_grid}

\begin{figure}[htbp]
\centering
\includegraphics[width=\columnwidth]{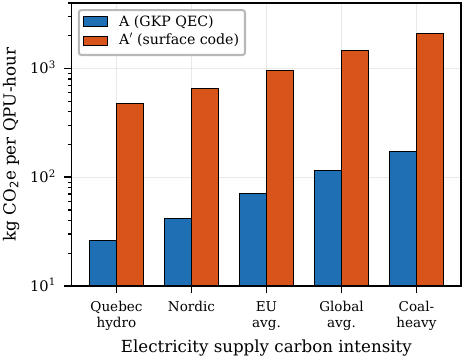}
\caption{Sensitivity of service-level carbon intensity to the carbon intensity of the electricity supply, at 50\% utilization and a five-year service life.}
\Description{Grouped bar chart on a logarithmic scale comparing two quantum scenarios across five electricity supply carbon intensities from hydroelectric to coal-heavy.}
\label{fig:grid}
\end{figure}

Retaining the grid carbon intensity $I(t)$ inside the operational integral of Equation~\ref{eq:op_integral} admits both siting and scheduling as levers. Figure~\ref{fig:grid} evaluates the service-level footprint across representative supply intensities.

Relocating Scenario~A from the assessed hydroelectric supply to a coal-heavy grid raises carbon intensity from 26.4 to 172.2\,kg\,CO\textsubscript{2}e per QPU-hour, a factor of 6.5. The corresponding factor for Scenario~A$'$ is 4.4. The sensitivity is weaker than utilization or service life precisely because the embodied term, which is invariant to the electricity supply, constitutes the majority of the footprint. Siting matters, but it cannot compensate for a poorly amortized manufacturing burden.

This ordering has a practical implication for providers. Where the assessed platform already operates on low-carbon electricity, as is the case for the inventory modeled here, further operational decarbonization offers limited headroom, and effort is better directed toward utilization and hardware retention. The result also qualifies the carbon-aware scheduling direction proposed in the CQC framework: for a platform whose power draw is 12.1:1 load-independent (Scenario~A), deferring jobs to low-intensity grid periods shifts only the marginal component, since the baseline draw continues regardless. Carbon-aware scheduling is therefore considerably less effective for cryogenic quantum platforms than for classical workloads, unless it is coupled with genuine consolidation that permits systems to be warmed and taken offline.

\subsection{Status of the Application-Centric Offset}
\label{sec:res_offset}

\begin{figure}[htbp]
\centering
\includegraphics[width=\columnwidth]{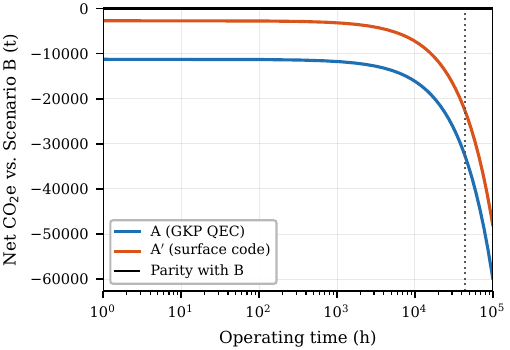}
\caption{Net life-cycle carbon difference between the quantum scenarios and classical Scenario~B. Curves below the parity line indicate that no application-centric offset is required for the quantum platform to be advantaged.}
\Description{Plot of net CO2-equivalent difference against operating time, with both quantum scenario curves lying entirely below the zero parity line.}
\label{fig:netdiff}
\end{figure}

Throughout the preceding analysis the application term of Equation~\ref{eq:cqc} has been held at $CO_2e_{\text{application}} = 0$, giving a conservative upper bound. Figure~\ref{fig:netdiff} reports the net life-cycle difference between each quantum scenario and the classical comparator, which determines whether that term needs to be invoked at all.

Both quantum scenarios lie below parity across the entire assessed range. At five years, Scenario~A is 32{,}506\,t\,CO\textsubscript{2}e below Scenario~B and Scenario~A$'$ is 22{,}520\,t below. The application term is therefore not required for either configuration to be environmentally preferable to the modeled classical comparator on a per-platform basis, and the conservative baseline is sufficient to support the comparison.

This does not diminish the term's importance. Its relevance is prospective rather than retrospective: it becomes decision-relevant when a provider must justify a platform whose embodied and operational burdens are high in absolute terms by reference to the downstream abatement its workloads enable. Populating Equation~\ref{eq:app_offset} requires a defined workload mix and domain-specific abatement factors, neither of which can be derived from a hardware inventory. We therefore retain it structurally, as the interface through which provider-specific workload data would enter the framework, and identify its quantification as the principal open problem in operationalizing CQC for cloud delivery.

\subsection{Synthesis for Quantum Cloud Service Providers}
\label{sec:res_synthesis}

Taken together, the results order the available levers by effect size. Utilization offers the largest response, at a factor of 19.7 across the assessed range, followed by service life extension at 59.9\% between three and ten years, and electricity supply at a factor of 6.5 in the worst case. All three act on terms that a provider controls directly, and none require modification to the underlying quantum hardware.

The ordering is itself the contribution. A provider reasoning by analogy from classical cloud infrastructure would prioritize operational efficiency and renewable procurement, which the classical comparators in Table~\ref{tab:decomp} would justify. Applied to an error-corrected superconducting platform, that ordering is close to inverted. The dominant levers are those that amortize a fixed manufacturing burden across more delivered computation, namely keeping the machine busy and keeping it in service.

\subsection{Threats to Validity}
\label{sec:res_threats}

Three limitations bound these results. First, the load-proportional fraction $\phi$ is not directly resolved by the inventory, which reports component power without distinguishing idle from active draw. We adopt $\phi = 0.20$ on physical grounds, since autonomous GKP correction is continuous, but the utilization finding strengthens as $\phi$ falls and weakens as it rises; at $\phi = 0.5$ the baseline ratio for Scenario~A falls from 12.1:1 to 4.2:1. Direct measurement of idle and active power on an operating platform would remove this assumption.

Second, service life is not modeled in the underlying inventory, and the per-job metrics depend on it linearly. We report results parametrically across one to fifteen years in Section~\ref{sec:res_life} rather than committing to a single figure, but component replacement within a service life, which would recur a portion of the embodied burden, is not captured.

Third, the functional unit is a platform-hour rather than an equivalent computational task. Because performance benchmarks for quantum computing remain unsettled, a comparison expressed per unit of useful computation is not currently constructible, a limitation the CQC framework itself identifies as an open challenge~\cite{Arora2025}. The comparison with the classical scenarios should accordingly be read as an infrastructure-level comparison at equal operating time, not as a claim of computational equivalence.

\section{Conclusion}
\label{sec:conclusion}

This paper extended the carbon-aware quantum computing framework of Arora and Kumar~\cite{Arora2025} to a service level and instantiated it with life cycle inventory data for a 100-logical-qubit superconducting platform, finding a five-year climate change burden of 583\,t\,CO\textsubscript{2}e (bosonic GKP) and 10{,}570\,t (surface-code), dominated by embodied carbon (77.3\% and 85.2\% respectively, chiefly cryostat production) with operational-embodied parity not reached until 17.0--28.7 years, well beyond any plausible service life and in sharp contrast to the 2.7-year parity of classical comparators. This embodied dominance reorders the sustainability levers available to QCS providers relative to classical-cloud intuition: utilization is the largest lever (a 19.7$\times$ reduction in carbon intensity from 5\% to full occupancy), followed by service life extension (59.9\% between three and ten years) and electricity supply (6.5$\times$ in the worst case), while operational efficiency, renewable procurement, and carbon-aware scheduling offer comparatively little leverage against a fixed, amortized manufacturing burden. These findings are subject to two boundaries: they reanalyze published inventory data rather than a measured operating facility, with the idle/active power split inferred rather than observed, and the platform-hour functional unit precludes a per-computation comparison against classical infrastructure given unsettled QC performance benchmarks. Future work should prioritize direct power measurement, populate the application-centric offset with provider-specific workload data, and extend the assessment to photonic, trapped-ion, and neutral-atom architectures to test whether embodied dominance generalises beyond superconducting cryogenics; until then, the practical implication for QCS providers is that the most effective sustainability measures are simply keeping a machine busy and keeping it in service.

\bibliographystyle{ACM-Reference-Format}
\bibliography{References}

\end{document}